\documentclass[a4paper,11pt]{article}

\usepackage{jheppub} 

\usepackage{epstopdf}
\usepackage{amsmath} 
\usepackage{amsthm} 

\usepackage{stmaryrd}

\title{Restoration of supersymmetry in  two-dimensional SYM with sixteen supercharges on the lattice}

\author[a]{Eric Gigu\`ere}
\author[b]{and Daisuke Kadoh}

\affiliation[a]{Department of Physics, University of Hokkaido, Sapporo, Hokkaido 060-0810, Japan}
\affiliation[b]{KEK Theory Center, High Energy Accelerator Research Organization (KEK),\\
            \ Tsukuba, Ibaraki 305-0801, Japan}

\emailAdd{eric@particle.sci.hokudai.ac.jp}
\emailAdd{kadoh@post.kek.jp}

\preprint{
{\flushright
 \vspace{-8mm}
KEK-TH-1797, EPHOU-15-004
\\
}}

\abstract
{
We perform lattice simulations of two-dimensional supersymmetric Yang-Mills theory with sixteen supercharges 
with a lattice action which has two exact supercharges (Sugino lattice action).
According to the gauge/gravity duality, the theory at finite temperature is expected to be well described 
by the corresponding black 1-branes, at low temperature in the large $N$ limit.
We aim to confirm the duality conjecture by comparing the lattice results
with the theoretical predictions obtained in the gravity side.
In this article,  at the beginning of this study, we examine the supersymmetric Ward-Takahashi identity
to test whether the lattice action reproduces the correct continuum theory.
Numerical results of the SUSY WTI strongly suggest us that any cut-off effects, 
which break supersymmetry, disappear in the continuum limit.
In addition, we study the issue of degenerate vacua and find that the admissiblilty condition 
or any other constraints of the link fields which guarantee the unique vacuum are not always needed.
}


\begin{document}
\maketitle

%
%
%
%
%

\section{Introduction}
\label{sec:intro}

The correspondence between gauge theory and the theory of gravity is an important subject in theoretical physics.
It was originally proposed as the AdS/CFT correspondence\cite{Maldacena:1997re} which was predicted in string theory.
Once we accept the duality, there are many interesting applications\cite{Klebanov:2000me}. 
The duality conjecture gives us deep insights into 
some long standing problems, such as the black hole information loss paradox\cite{Hawking:1976ra}. 
Moreover, there are many attempts to solve strongly coupled theories, like QCD, analytically in the gravity side\cite{Sakai:2004cn}, \cite{Herzog:2009xv}.
To ensure the validity of the applications, it is important to give strong evidence of the duality conjecture.
In this study, we aim to verify it by using lattice simulations of supersymmetric Yang-Mills theories.

Lattice theory provides a framework to study strongly coupled gauge theories. 
In the context of the gauge/gravity duality, gauge theory is typically a strongly coupled one and thus is a good target for lattice theory.
We can test the duality conjecture by comparing lattice results in the gauge theory side with theoretical predictions in the gravity side.
In principle, unlike localization methods\cite{Pestun:2007rz} and other analytical treatments\cite{Erickson:2000af} for BPS states, 
 lattice gauge theory allows us to test it for any physical quantities and independently of $N$ and 't  Hooft coupling constant $\lambda$.
Although lattice theory achieves great success for Yang-Mills theories, the nature of supersymmetry algebra makes it hard to put supersymmetric Yang-Mills theories on the lattice. 
Therefore, to test the gauge/gravity duality for SUSY theories using lattice methods has been a difficult issue for a long time.

In the last decade,
some developments of lattice supersymmetry changed the situation completely.
Several lattice formulations of SYM theories were proposed in
 \cite{Kaplan:2002wv}, 
\cite{Sugino:2003yb}, \cite{Sugino:2004qd}, \cite{Sugino:2004uv}, \cite{Sugino:2006uf}, 
\cite{Matsuura:2014pua},
\cite{Catterall:2004np}, 
\cite{D'Adda:2005zk},
\cite{Matsuura:2008cfa},
\cite{Joseph:2013jya} 
and enabled us to perform direct tests of the duality. 
In one dimensional SYM with 16 supercharges, 
some evidences of the correspondence between the gauge theory and the dual black hole\cite{Itzhaki:1998dd} have been obtained from numerical simulations\cite{Anagnostopoulos:2007fw},
\cite{Catterall:2007fp}. 
In two-dimensional ${\cal N}=(8,8)$ SYM, lattice simulations have been already performed and the phase structure of the theory was investigated\cite{Catterall:2010ya}.  
However, this subject deserves more studies 
because some physical quantities, like the internal energy of the  black 1-branes\cite{Klebanov:1996un}, have not been estimated yet.
Therefore, we perform lattice simulations of ${\cal N}=(8,8)$ SYM in two dimensions by employing the Sugino lattice action\cite{Sugino:2004uv}.

In this paper,  as the first step of our study, we examine the behavior of the Sugino lattice action for  ${\cal N}=(8,8)$ supersymmetric Yang-Mills theory in two dimensions.
By examining the SUSY Ward-Takahashi identity, we confirm that supersymmetry is restored  in the continuum limit.
This paper is organized as follows: We briefly review the Sugino lattice model and discuss degenerate vacua in section 2. 
Then, we see numerical results of the SUSY Ward-Takahashi identity in section 3. 
The last section 4 is devoted to a summary of our results.
In appendix A,  we give a relation between the notations used in this paper.  The full lattice action can be found  in appendix B.

%
%
%
%
%

\section{Two-dimensional ${\cal N}=(8,8)$ SYM}
\label{sec:swti}

In this section,  we explain the continuum action of the target theory and the lattice action we employed. 
We use Sugino's method to define the lattice action\cite{Sugino:2004uv}. After introducing the continuum action we rewrite it as a $Q_\pm$-exact form, then introduce the lattice action from the $Q_\pm$-exact one.
This way, the supersymmetries associated with the two supercharges $Q_\pm$ are kept exactly on the lattice.

%
%

\subsection{Continuum theory}

Two-dimensional supersymmetric Yang-Mills theory with sixteen supercharges contains two gauge fields $A_\mu(\mu=0,1)$, 
and eight scalar fields $X_i(i=2,3\cdots,9)$, and sixteen fermions $\Psi_\alpha(\alpha=1,2,\cdots,16)$ as the fundamental field variables. 
The scalars and fermions, as well as the gauge fields, are in adjoint representation of $SU(N)$ and can be represented as matrix-valued fields. \footnote{
Let $\varphi$ be a scalar field, a gauge field, or a fermi field.
In this paper, $\varphi$ can be expanded as
$\varphi=\sum _a \varphi^a T^a$ where $T^a$ are group generators satisfying ${\rm tr}(T^aT^b)=\delta_{ab}$,
}

The Euclidean action is given by
\begin{eqnarray}
&& 
       S=\frac{N}{\lambda}\int{}{\rm d}^2x\ {\rm tr} 
             \left\{ \frac{1}{4}F^2_{\mu\nu} 
                    + \frac{1}{2}(D_\mu X_i)^2 - \frac{1}{4}[X_i,X_j]^2  
             \right.  \nonumber \\
&& 
      \hspace{3cm} \left.
          +\frac{1}{2} \Psi_\alpha  D_0 \Psi_\alpha 
          -\frac{i}{2} \Psi_\alpha (\gamma_{1})_{\alpha \beta} D_1 \Psi_\beta 
        + \frac{1}{2}\Psi_\alpha (\gamma_i)_{\alpha \beta} [X_i, \Psi_\beta] \right\},
        \label{cont_action}
\end{eqnarray}
where, unless we use twisted fields which are defined later,
the Roman indices $i,j$ run from $2$ to $9$, 
the Greek indices associated with the space-time directions $\mu,\nu$ take the value $0$ or $1$,
while the Greek indices  associated with spinors $\alpha, \beta$ run from $1$ to $16$. 
The field strength is $F_{01} = \partial_0 A_1 - \partial_1 A_0 +i[A_0,A_1]$ and the covariant derivatives are 
defined by $D_\mu \varphi = \partial_\mu \varphi + i[A_\mu, \varphi]$.  
In addition, $\gamma_a(a=1,\cdots,9)$ are real symmetric matrices that satisfy the nine-dimensional Euclidean Clifford algebra,
$\{\gamma_a,\gamma_b\}=2\delta_{ab}$.

This action is invariant under the following sixteen supersymmetry transformations,
\begin{eqnarray}
&& Q_\alpha A_\mu = -i (\gamma_\mu)_{\alpha\beta} \Psi_\beta,  \label{Q_cont_A}\\ 
&& Q_\alpha X_i = -i  (\gamma_i)_{\alpha\beta} \Psi_\beta, \label{Q_cont_X} \\
&& Q_\alpha \Psi_\beta =  i(\gamma_1)_{\alpha\beta} F_{01} + (\gamma_\mu\gamma_i)_{\alpha\beta} D_\mu X_i
  +\frac{i}{2} (\gamma_i\gamma_j)_{\alpha\beta} [X_i,X_j], \label{Q_cont_psi}
\end{eqnarray}
where $\gamma_0=i$.
Furthermore, the two-dimensional Poincar\'e group and the internal $O(8)$ rotations are symmetries of this theory.

The easiest way to give the action (\ref{cont_action}) is to perform a dimensional reduction from ten-dimensional ${\cal N}=1$ SYM.
However instead of going directly to two dimensions, reducing the theory to four dimensions first, 
we obtain four-dimensional ${\cal N}=4$ SYM which have a known topologically twisted version\cite{Dijkgraaf:1996tz}.
In the topological field theory, two of the supercharges $Q_\pm$ are scalars and nilpotent.
Its action is written in a $Q_\pm$-exact form which is needed for the lattice action defined in the next section. 
Then, by reducing dimensions from four to two, we can give another representation of the action (\ref{cont_action}).
\footnote{See appendix A for the relation between both notations.}

As a result, the action (\ref{cont_action}) can be written in the $Q_\pm$-exact form with the four-dimensional topological field theory notation,
\begin{eqnarray}
&& S = Q_+Q_-\frac{N}{2\lambda}\int{}{\rm d}^2x\ {\rm tr}  
   \left\{  - 4i B_i F_{i3}^+ 
             - \frac{2}{3} \epsilon_{ijk} B_i B_j B_k
    \right.\nonumber \\
&& \hspace{4.5cm} \left. 
           - \psi_{+\mu}\psi_{-\mu}
           - \chi_{+i}\chi_{-i}
           - \frac{1}{4}\eta_+\eta_- \right\},
           \label{exact_cont_action}
\end{eqnarray}
where  $\mu$ runs from $0$ to $3$, and $i,j,k$ from $0$ to $2$, and  $\epsilon_{ijk}$ is a totally antisymmetric tensor that satisfy $\epsilon_{012}=1$.
Note that $\mu$ and $i$ are different from those of the original variables, $A_\mu,X_i$.
Here, the bosonic fields consist of two gauge fields $A_0, A_1$, six real scalar fields  $A_2, A_3, B_i, C$, and two complex scalar fields $\phi_\pm$.
The fermions are given by $\psi_{\pm \mu},\ \chi_{\pm i},\ \eta_\pm$.
The extended field strengths are 
\begin{eqnarray}
&& F_{i3}^+ = \frac{1}{2} \left (F_{i3} + \frac{1}{2}\epsilon_{ijk}F_{jk} \right),\\
&& F_{\mu \nu}= \partial_\mu A_\nu - \partial_\nu A_\mu +i[A_\mu,A_\nu], \label{Fmunu}
\end{eqnarray}
with $\partial_\mu =0$ for $\mu=2,3$.

The associated $Q_\pm$-transformation are
\begin{center}
\begin{tabular}{ l  l  l }
$Q_\pm A_\mu = \psi_{\pm \mu} $     & $Q_\pm \psi_{\pm \mu} = -iD_\mu \phi_\pm,  $  & $Q_\pm \psi_{\mp \mu} = \frac{i}{2} D_\mu C \pm \tilde{H}_\mu, $\\[5pt]
$Q_\pm B_i = \chi_{\pm i},    $     & $Q_\pm \chi_{\pm i}   = [B_i,\phi_\pm],    $  & $Q_\pm \chi_{\mp i} = \frac{1}{2} [C,B_i] \pm H_i,             $\\[5pt]
$Q_\pm C = \eta_\pm,          $     & $Q_\pm \eta_\pm = [C,\phi_\pm],           $  & $Q_\pm \eta_\mp = [\phi_\mp,\phi_\pm],                         $\\[5pt]
$Q_\pm \phi_\pm = 0,          $     & $Q_\pm \phi_\mp = \eta_\mp,               $  & 
\end{tabular}
\begin{align}
&Q_\pm H_i \,        = \pm\Big(  [\chi_{\mp i},   \phi_\pm]  + \frac{1}{2}[\chi_{\pm  i }, C] + \frac{1}{2}[B_ i, \eta_\pm  ] \Big),\nonumber \\
&Q_\pm \tilde{H}_\mu = \pm\Big(  [\psi_{\mp \mu}, \phi_\pm]  + \frac{1}{2}[\psi_{\pm \mu}, C] - \frac{i}{2}D_\mu \eta_\pm \Big).
\label{def:Q-transf}
\end{align}
\end{center}
We introduced seven auxiliary fields $\tilde{H}_\mu$ and $H_i$ to define $Q_\pm$ as closed transformations.
With this arrangement, the $Q_\pm$-transformations are nilpotent up to gauge transformations, 
$Q_\pm^2 = i \delta_{\phi_\pm}$ and $\{Q_+,Q_-\} = -i\delta_C$ with $\delta_\omega$ being a gauge transformation with a gauge parameter $\omega$, 
\begin{eqnarray}
\delta_\omega A_\mu=-D_\mu \omega, \qquad \delta_\omega \varphi = -i[\varphi,\omega], 
\end{eqnarray}
where $\varphi$ is either a scalar or a  fermion.

%
%

\subsection{Lattice theory}

We consider the lattice theory on a two-dimensional lattice with a lattice space $a$, and hereafter we set $a=1$ for simplicity.
Let $N_t,N_x$ be the temporal and spatial lattice sizes, respectively. Then, 
the lattice sites $\vec{x}=(t,x)$ are labeled by integers,
\begin{eqnarray}
t=1, \cdots, N_t, \qquad  x=1,\cdots, N_x.
\end{eqnarray}
The gauge fields are defined on the links
as gauge group-valued fields $U_\mu(\vec{x})$,
and the other fields are defined on the sites.
In this subsection, we impose periodic boundary conditions on every field for both temporal and spatial directions in order to discuss SUSY invariance of the lattice action. In section \ref{sec:susy},
the fermions satisfy anti-periodic boundary conditions for the temporal direction 
because we perform numerical simulations at finite temperature.

The infinitesimal lattice gauge transformations for the link fields and the other adjoint fields $\varphi$ are given by
\begin{eqnarray}
\delta_\omega U_\mu (\vec{x}) =i\omega(\vec{x}) U_\mu (\vec{x}) - i U_\mu (\vec{x}) \omega(\vec{x}+\hat\mu), \qquad \ \ 
 \delta \varphi(\vec{x}) = -i [\varphi(\vec{x}), \omega(\vec{x})],
\label{lat_gauge_transf}
\end{eqnarray}
where $\hat \mu$ is a unit vector in $\mu$-direction and $\omega$ is a gauge parameter defined on the sites. 
The forward and backward covariant difference operators are given by
\begin{eqnarray}
&& \nabla^+_\mu \varphi(\vec{x}) = U_\mu(\vec{x})                  \varphi(\vec{x}+\hat{\mu})U_\mu^\dagger(\vec{x}) - \varphi(\vec{x}), 
\label{fwd_diff_op}
\\
&& \nabla^-_\mu \varphi(\vec{x}) = \varphi(\vec{x}) - U_\mu^\dagger(\vec{x}-\hat{\mu})\varphi(\vec{x}-\hat{\mu})U_\mu(\vec{x}-\hat{\mu}),
\end{eqnarray}
respectively.

The use of the link fields leads to extra difficulties and forces us to adapt the lattice counterparts of the $Q_\pm$-transformations in order to assure that they are nilpotent and to keep gauge invariance intact. In fact, 
some modifications are needed only for the fields with directions $\mu=0,1$.
The modified $Q_\pm$-transformations for them are
\begin{align}
\begin{split}
& Q_\pm U_\mu = i\psi_{\pm \mu}U_\mu,  \\
& Q_\pm \psi_{\pm \mu} = - i\nabla^+_\mu \phi_\pm +  i\psi_{\pm \mu}\psi_{\pm \mu},\\
& Q_\pm \psi_{\mp \mu} = \frac{i}{2} \nabla^+_\mu C \pm \tilde{H}_\mu+\frac{i}{2}\{\psi_{+\mu}, \psi_{-\mu} \}, \\
& Q_\pm \tilde{H}_\mu = \pm \Big(  \left [ \psi_{\mp \mu}, \phi_\pm+ \tfrac{1}{2} \nabla_\mu^+ \phi_\pm \right ]
+ \frac{1}{2} \left [\psi_{\pm \mu},        C+       \tfrac{1}{2} \nabla_\mu^+ C \right] \\
 &\hspace{2cm}- \frac{i}{2}\nabla^+_\mu \eta_\pm
+ \frac{1}{4} [\psi_{\pm \mu}, \{ \psi_{+\mu}, \psi_{-\mu} \}  \pm 2i\tilde{H}_\mu] \Big).
\label{Lat_Q}
\end{split}
\end{align}
While the transformation law for the other fields is the same as the continuum one.
From the definitions above, the $Q_\pm$-transformations are nilpotent up to the lattice gauge transformations (\ref{lat_gauge_transf}). 
We have $Q_\pm^2 = i\delta_{\phi_\pm}$ and $\{Q_+,Q_-\} = -i\delta_C$ even on the lattice\cite{Sugino:2004uv}.

Let us give the lattice action from (\ref{exact_cont_action}) by replacing the integral with a summation over the sites
and changing the field strengths (\ref{Fmunu}) to lattice counterparts,
\begin{eqnarray}
&& S = Q_+Q_-\frac{N}{2\lambda_0}\sum_{t,x} \ {\rm tr}  
   \left\{  - 4i B_i F_{i3}^+ 
             - \frac{2}{3} \epsilon_{ijk} B_i B_j B_k
    \right.\nonumber \\
&& \hspace{4.5cm} \left. 
           - \psi_{+\mu}\psi_{-\mu}
           - \chi_{+i}\chi_{-i}
           - \frac{1}{4}\eta_+\eta_- \right\},
           \label{lat_action}
\end{eqnarray}
where the extended field strengths must be carefully chosen,
\begin{align}
F_{03}^+ &= \tfrac{1}{2}(\nabla^+_0   A_3 + \nabla^+_1   A_2 ), \label{Fp01} \\
F_{13}^+ &= \tfrac{1}{2}(\nabla^-_1 A_3 - \nabla^-_0 A_2 ), \label{Fp02} \\
F_{23}^+ &= \tfrac{1}{2}(i[A_2 ,A_3 ] + F_{01}),  \label{Fp03}
\end{align}
with
\begin{align}
F_{01}(\vec{x})&=-\frac{i}{2}\left(P_{01}(\vec{x}) - P_{01}^\dagger(\vec{x}) - \frac{1}{N} {\rm tr}(P_{01}(\vec{x}) - P_{01}^\dagger(\vec{x}))\right),
\label{F01}
\\
P_{01}(\vec{x})&=U_0(\vec{x})U_1(\vec{x}+\hat{0})U_0^\dagger(\vec{x}+\hat{1})U_1^\dagger(\vec{x}).
\label{P01}
\end{align}
See \cite{Sugino:2004uv} for more details.
If the forward-type difference operators are used for both (\ref{Fp01}) and  (\ref{Fp02}), 
then the lattice action has fermionic and bosonic doublers. This mixed choice forbids the doublers.
The hermiticity of $F_{01}$ leads to a semi positive definite boson action. 
The trace-part of $F_{01}$ makes no contribution to the lattice action.
Therefore we employed the traceless hermitian $F_{01}$ (\ref{F01}).
Moreover, by performing the $Q_\pm$-transformation (\ref{Lat_Q}), a concrete form of the lattice action can be obtained. 
However, it is somewhat complex and lengthy, therefore we give it in appendix B.

Note that,  our simple choice of $F_{01}$  (\ref{F01}) yields a problem of extra vacua,
as was first pointed out in \cite{Sugino:2003yb}  and detailed in the next section. To avoid the problem, we can take 
an admissibility-type field tensor\cite{Sugino:2004qd} or a ${\rm tan}(\theta/2)$-type field tensor\cite{Matsuura:2014pua}. To use them is theoretically clear because the extra vacua are forbidden.
But to write a simulation code using such field tensors is complex and demanding task, therefore we chose the simple field tensor (\ref{F01}).
As seen in section \ref{subsec:results}, no transitions from the desirable vacuum to the extra vacua are observed. Therefore, to use  (\ref{F01}) is not a problem in practice, at least, for simulation parameters on which we focus in this paper.

%
%

\subsection{Continuum limit and degenerate vacua}
\label{subsec:many_vacua}
In this section, we discuss the naive continuum limit of the lattice action and the presence of degenerate vacua\cite{Sugino:2003yb}.
We argue that the degenerate vacua, while making the theoretical study of the continuum limit difficult, do not pose any serious problems in the simulations when testing the duality conjecture.

Let us consider the  naive continuum limit of the lattice action (\ref{lat_action}).
The 't Hooft coupling $\lambda$ has mass dimension two which gives the typical scale of this theory.
Therefore, $a \rightarrow 0$ is realized by taking $\lambda_0(=\lambda a^2) \rightarrow 0$ with fixed $\lambda$.
Then, dimensions of the fields indicate
\begin{eqnarray}
U_\mu^{lat.} =1 +iaA^{cont.}_\mu+\cdots,\qquad 
X^{lat.}_i  =aX_i^{cont.}, \qquad \Psi^{lat.}  = a^{3/2} \Psi^{cont.},
\label{lattice_to_cont}
\end{eqnarray}
where the fields with $lat.$ are dimensionless fields which define the lattice action (\ref{lat_action}), 
\footnote{
We can express the lattice action (\ref{lat_action}) in the original variables $A_\mu,X_i,\Psi_\alpha$  by using the relations given  in appendix A. 
}
while the fields with $cont.$ are the ones in the continuum action (\ref{cont_action}). 
After performing the $Q_\pm$-transformations,  the lattice action has many higher derivative terms. 
By neglecting the higher derivative terms and using (\ref{lattice_to_cont}),  we find that, in the continuum limit,
  the lattice $Q_\pm$-transformations coincide with the continuum ones and the lattice action (\ref{lat_action}) reproduces the correct continuum action  (\ref{cont_action}).

This argument of the continuum limit can be justified if
\begin{eqnarray}
P_{01}(\vec{x}) \rightarrow 1, \qquad C \rightarrow 0,  \qquad {\rm as} \ \ \lambda_0 \rightarrow 0,
\label{conditions_continuum_limit}
\end{eqnarray}
where $P_{01}$ are the plaquettes (\ref{P01}) and $C$ is the sum of the higher derivative terms 
in the bosonic
action.\footnote{ The higher derivative terms in the bosonic sector are given by
\begin{eqnarray}
C = \frac{N}{\lambda_0} \sum_{t,x} {\rm tr} \left\{[X_2 ,X_3 ][X_5,X_6] + [X_2, X_5^{+0}][X_6, X_3^{+0}] + [X_2^{+1},X_6][X_3, X_5^{+1}] + \cdots \right\},
\label{C_term}
\end{eqnarray}
where $ X_i^{+\mu} = X_i + \nabla^+_\mu X_i$.
In the naive continuum limit, the three terms above cancel each other by the Jacobi identity. The complete equation of $C$ is given in (\ref{full_c_term}). 
}
When $P_{01} \rightarrow 1$,   the link fields also approach unity up to gauge transformations, therefore the expansion around one $U_\mu=1+iaA_\mu + \cdots$ is justified. 
\footnote{
Here, we assume that the lattice size is infinite.
}
We only have to study the higher derivative terms $C$ in the bosonic sector
because, when $C$ vanishes, the corresponding higher derivative terms in the fermionic sector also vanish thanks to  the $Q_\pm$-symmetries.

We have to examine (\ref{conditions_continuum_limit})  numerically for each simulation parameter
because it does not hold \textit{in  general} as explained below.
When the scalar fields are set to zero, the boson action is given by the  so-called double plaquette action,
\begin{eqnarray}
\left. S_{B} \right|_{X_i=0} = \frac{N}{2\lambda_0} \sum_{t,x} {\rm tr} \left\{-\frac{1}{4} \left(P_{01}-P_{01}^\dag
- \frac{1}{N} {\rm tr}(P_{01} - P_{01}^\dag)
\right)^2 \right\}.
\label{double_plaquette_action}
\end{eqnarray}
We find that $\lambda_0 \rightarrow 0$ does not immediately mean $P_{01}=1$. 
For example in the SU(2) case, $P_{01}=-1$  is  another possibility which gives a zero point of the integrand in (\ref{double_plaquette_action}).
For non-zero scalar fields, the situation becomes more serious. 
For the simplicity of the explanation, let $X_2,X_3$ be non-zero and the other scalars be zero.  
Then,  the boson action can be given as follows:
\begin{eqnarray}
\left. S_{B}\right|_{X_4=\cdots=X_9=0} = \frac{N}{2\lambda_0} \sum_{t,x} \sum_{i=0}^2{\rm tr} \left\{ \varphi_i^2   \right\},
\label{lattice_boson_action_e}
\end{eqnarray}
where
\begin{eqnarray}
&&\varphi_0 = \nabla^+_0   X_3 + \nabla^+_1   X_2,   \\
&&\varphi_1 = \nabla^-_1 X_3 - \nabla^-_0 X_2,       \\
&& \varphi_2 = i[X_2 ,X_3 ]   + F_{01}.
\end{eqnarray}
Near the continuum limit, if $\nabla_\mu X_i \sim 0$, that is, if $X_i$ are smooth, then the higher derivative terms actually disappear. 
However, the conditions $\varphi_i=0$ imply that there can exist extra minima other than $\nabla_\mu X_i =0$.
For example, non-smooth configurations that satisfy  $\nabla^+_0 X_3 =  -\nabla^+_1 X_2={\cal O}(1)$ give $\varphi_0=0$. 
Therefore, $C$ could survive as $\lambda_0 \rightarrow 0$.

Fortunately, the condition  (\ref{conditions_continuum_limit}) is likely to hold for parameters in which we are interested.
We actually consider the theory at finite temperature and test the gauge/gravity duality in the large $N$ limit.
We expect that the undesirable configurations are suppressed by temperature effects or by the influence of the large $N$ limit.
For the gauge fields, the two minima $P_{01}=\pm1$ are separated by a finite barrier associated with the saddle point of (\ref{double_plaquette_action}).
The height of the barrier becomes larger in the large $N$ limit or in the continuum limit
because the action has $N/\lambda_0$ as an overall factor.
Furthermore,  for the scalar fields, a potential reducing the fluctuations of the fields is dynamically generated by SUSY breaking temperature effects.
At high temperature, the potential makes the scalar fields $X_i$ stay near zero.
These suppress transitions from the trivial vacuum $U_\mu=1$ and $X_i=0$ to the others, for example, from $P_{01}=1$ to $P_{01}=-1$ etc.
Therefore, it is reasonable to consider that the lattice action with the naive field tensor (\ref{lat_action}) is sufficient to test the duality conjecture,  despite the need of numerical confirmations of  (\ref{conditions_continuum_limit}).
\footnote{
As seen in section 3.3, 
starting the HMC from a trivial configuration $U_\mu=1$ and $X_i=0$ at the temperature $T/\lambda^{1/2}=0.3$ in $SU(2)$,
we observed that the higher derivative terms $C$ were small in the run. Moreover, $P_{01}$ tended to stay close to unity. 
In this case, we added a mass term for the scalar fields to the action. The mass term forbids large fluctuations of the scalar fields and contributes to the realization of $C \rightarrow 0$ as $\lambda_0 \rightarrow 0$ for such low temperature and small $N$.  While, somewhat surprisingly, $P_{01} \simeq -1$ was not observed for the simulation parameters even for $a\lambda^{1/2} \simeq 0.28$. We have considered that the naive choice of the field tensor (\ref{F01}) works at high temperature, in the large N, or near the continuum limit. Contrary to our guess, it may work for a wider range of the simulation parameters.
}


\section{Restoration of supersymmetry}
\label{sec:susy}

In this section we verify the restoration of the full supersymmetry in the continuum limit.
The present lattice model breaks fourteen of the sixteen supersymmetries at a finite lattice spacing.
At least, in the perturbation theory,  keeping partial supercharges on the lattice guarantees the restoration of the full symmetry without the usage of any fine tunings.
We numerically estimate the supersymmetric Ward-Takahashi identity to verify the SUSY restoration beyond the perturbation theory. \footnote{ See \cite{Kanamori:2008bk} for ${\cal N}=(2,2)$ SYM.}

%
%

\subsection{Simulation details}
\label{subsec:simulations}

Numerical simulations are made using the rational Hybrid Monte Carlo method\cite{Clark:2004cp}.
As detailed below, the field configuration of the HMC consists of twelve bosons: two gauge fields, eight scalar fields and two auxiliary fields.
Moreover, the dynamical effects of the fermions are treated using the pseudo-fermion method with the rational approximation with the exception of the phase of the pfaffian. 
The effects of the phase are included in the numerical results of the SUSY WTI presented in this paper, using the phase reweighting method.

The present lattice action has two four-fermi interactions,
\begin{eqnarray}
S_{4f} =\frac{N}{2\lambda_0} \sum_{t,x} \sum_{\rho=0,1}{\rm tr}\left( -\frac{1}{4}\{\psi_{+ \rho},\psi_{-\rho}\}^2 \right),
\label{four_fermi}
\end{eqnarray}
 which are of the order of the cut-off.
These terms originally come from the two fermi terms in the right-hand side of the supertransformation of the link fields (\ref{Lat_Q}).
If we truncate these terms, the lattice action loses the two exact supercharges that are keys in the supersymmery restoration.
Therefore, the terms (\ref{four_fermi}) should be included in the simulations.
By introducing two auxiliary fields, we can express (\ref{four_fermi}) as 
\begin{eqnarray}
S_{4f} = \frac{N}{2\lambda_0}  \sum_{t,x} \sum_{\rho=0,1}{\rm tr}\left( \sigma_\rho^2 + \psi_{+ \rho} [\sigma_\rho, \psi_{-\rho}]  \right).
\label{auxiliary_field_action}
\end{eqnarray}
Hence, the fermion action is written in a bilinear form which is suited to the simulations.

Then, to integrate fermions gives the pfaffian,  
\begin{eqnarray}
{\rm pf} (D) = \int {\rm D} \Psi\, {\rm e}^{-\Psi^T D\Psi},
\end{eqnarray}
which is generally complex due to the Majorana-Weyl nature of the ten-dimensional fermions. 
We deal with the absolute value and the phase of the pfaffian individually.
Since ${\rm det}(D) = {\rm pf} (D)^2$,
the absolute value of the pfaffian can be prepared by a complex pseudo fermion $\phi$ as
\begin{eqnarray}
|{\rm pf}(D)| &=& {\rm det}(D^\dag D)^{1/4},\\
 &=& \int D\phi^\dag D \phi \ {\rm exp}\left\{
- \phi^\dag (D^\dag D)^{-\frac{1}{4}} \phi
\right\},
\end{eqnarray}
with the rational approximation of the fourth-root of $D^\dag D$,
\begin{eqnarray}
(D^\dag D)^{-1/4} \simeq \alpha_0 + \sum_{i=1}^{N_r} \left(  \frac{\alpha_i}{D^\dag D + \beta_i}\right).
\end{eqnarray}
To compute the inversions of $D^\dag D+\beta_i$, we use the multiple shift CG solver\cite{Frommer:1995ik}. 
The degree $N_r$ and the coefficients $\alpha_i,\beta_i$ depend on the accuracy of the approximation
and the maximal/minimum eigenvalues of $D^\dag D$ \cite{Clark:2005int}. 
They are determined from the eigenvalues at the thermalization step and fixed thereafter.
Moreover, we compute the phase of the pfaffian for every configurations and
use the phase reweighting method to obtain results involving the whole dynamical effects of the fermions.

The classical bosonic action in (\ref{cont_action}) has the following flat directions,
\begin{eqnarray}
A_\mu=0,   \quad X_i= {\rm constant \  and \ diagonal \ matrices}.
\end{eqnarray}
At high temperature $T_{\rm eff}={T/\lambda^{1/2}} \gg 1$, 
the HMC stably runs with small fluctuations of $X_i$ around zero and $P_{01}$ around unity.
However, at low temperature $T_{\rm eff} \lesssim 1$,
the HMC can get unstable by entering a flat direction.
Then, the magnitude of the scalar fields $R=\sum {\rm tr}( X_i^2)$ monotonically increases.
To avoid the instability,
we added a supersymmetry breaking mass term\cite{Kanamori:2008bk},
\begin{align}
S_{mass}=\frac{N}{2\lambda_0}\sum_{t,x}\sum_{i=2}^{9} \, m^2_0 \  {\rm tr} (X_i^2),  
\label{mass_term}
\end{align}
to the action, where $m_0=ma$ is the mass in the lattice unit. This controllable supersymmetry breaking term enable us 
to investigate the SUSY breaking cut-off effects by examining the SUSY WTI and to discuss the restoration of supersymmetry in the continuum limit.

The trajectory length is $0.5$ and the time step of the molecular dynamics is chosen to keep the acceptance rate over 80\%.
We take $N_r,\alpha_i, \beta_i$ of the rational approximation to guarantee an accuracy of $10^{-13}$ 
and compute the maximum and minimum eigenvalues every trajectories for confirmation.

%
%

\subsection{SUSY Ward-Takahashi identity}
\label{subsec:swti}

In the present model, supersymmetry is broken by three sources: finite temperature, the mass of the scalars and the lattice spacing.
The breaking effects of the temperature and the mass are physical, while the effect of the finite lattice spacing is purely a lattice artifact. 
We have to show that the lattice artifact vanishes in the continuum limit.  
As explained below, 
to observe the symmetry restoration, we can use the Ward-Takahashi identity including the breaking effect of the mass term (partially conserved supercurrent)\cite{Kanamori:2008bk}.

If the action is invariant under supersymmetry, then
the SUSY Ward-Takahashi identity holds as a manifestation of the symmetry.
At finite temperature, supersymmetry is broken due to the fermions obeying anti-periodic boundary conditions. 
However we can formally derive the same Ward-Takahashi identity using the path integral formulation.
This enables us to use the SUSY WTI to examine the behavior of the supersymmetry breaking effect which comes from the lattice spacing at finite temperature\cite{Kanamori:2008bk}.

In the continuum theory, the SUSY WTI is given by
\begin{align}
\partial_\mu J_{\mu}( \vec{x})  = \frac{m^2}{\lambda} Y( \vec{x}),
\label{SWTI}
\end{align}
where $\mu=0,1$,  and the supercurrents are
\begin{eqnarray}
J_{\mu}= -\frac{N}{\lambda} 
                      \left\{  \gamma_0 \gamma_1 \gamma_\mu {\rm tr} (\Psi F_{01}) 
                                                         + \gamma_\nu \gamma_i \gamma_\mu   {\rm tr} (\Psi D_\nu X_i) 
                                                          +\frac{i}{2} \gamma_i \gamma_j \gamma_\mu   {\rm tr} (\Psi [X_i,X_j])
                      \right\}.
\label{supercurrent}
\end{eqnarray}
The supersymmetry breaking mass term gives  
\begin{align}
Y = N  \gamma_i {\rm tr}(X_i \Psi),
\end{align}
which is the right-hand side of (\ref{SWTI}).
Both the supercurrents $J_\mu$ and the breaking term $Y$ are fermionic operators.

Let us consider the SUSY WTI as a correlation function because the fermionic one point function $\langle J \rangle $ identically vanishes. 
We choose, for simplicity, $Y_\beta(y)$ as a source operator. Then, 
the resultant WTI is
\begin{align}
 \partial_\mu \langle J_{\mu,\alpha}( \vec{x}) Y_\beta( \vec{y} )\rangle  = \frac{m^2}{\lambda} \langle Y_\alpha( \vec{x})Y_\beta(\vec{y}) \rangle - \delta^{2}( \vec{x}-\vec{y}) \langle Q_\alpha Y_\beta( \vec{y} )\rangle,
 \label{SWTI_as_correlator}
\end{align}
where $Q_\alpha$ are the supercharges. 
If the SUSY WTI holds, the ratios
\begin{align}
\frac{  \partial_{\mu} \langle J_{\mu,\alpha}( \vec{x}) Y_\beta( \vec{y} )\rangle}{ \langle Y_\alpha( \vec{x})Y_\beta( \vec{y} )\rangle}
\label{eq:ra}
\end{align}
should give the dimensionless mass ${m^2/\lambda}$ everywhere except at $ \vec{x}=\vec{y}$.
Therefore, by investigating lattice counterparts of the ratios,  
we can extract the supersymmetry breaking effect that comes from the lattice spacing.

On the lattice, the supercurrent is prepared using only the forward-type operators.
We use (\ref{F01}) for the field strength $F_{01}$ of (\ref{supercurrent}), and (\ref{fwd_diff_op}) for the covariant derivative of (\ref{supercurrent}).
Meanwhile, 
we use the naive symmetric difference operator when calculating the gradient of the supercurrent,
\begin{eqnarray}
\partial_\mu^s J_\mu( \vec{x})
 = \sum_{\mu=0,1}\frac{ J_\mu ( \vec{x}+\hat \mu ) - J_\mu ( \vec{x}-\hat \mu )} {2}.
\end{eqnarray}

%
%
%
%
%

\subsection{Numerical results}
\label{subsec:results}

\begin{table}
\begin{center}
\begin{tabular}{| c c c c |}
  \hline                       
  $\ \ N_t$&$N_x$ & $ m^2/\lambda$ & Nconf \\
  \hline                   
  12 &  6 & 1.00 & 2000 \\
  12 &  6 & 0.75 & 2000 \\
  12 &  6 & 0.50 & 2000 \\
  12 &  6 & 0.25 & 2000 \\
  \hline                   
  16 & 8  & 1.00 & 2000 \\  
  16 & 8  & 0.75 & 2000 \\  
  16 & 8  & 0.50 & 2000 \\  
  16 & 8  & 0.25 & 2000 \\
  \hline                   
  20 & 10 & 1.00 & 2000 \\
  20 & 10 & 0.75 & 2000 \\
  20 & 10 & 0.50 & 2000 \\
  20 & 10 & 0.25 & 2000 \\
  \hline  
\end{tabular}
\caption{Simulation parameters and the number of configurations. The dimensionless temperature is $T_{\rm eff}=0.3$ and the lattice spacing is determined by $a \lambda^{1/2}=1/(N_tT_{\rm eff})$.}
\label{tab:para}
\end{center}
\end {table}

\begin{figure}[t]
\begin{center}
 {\includegraphics[width=0.85\columnwidth]{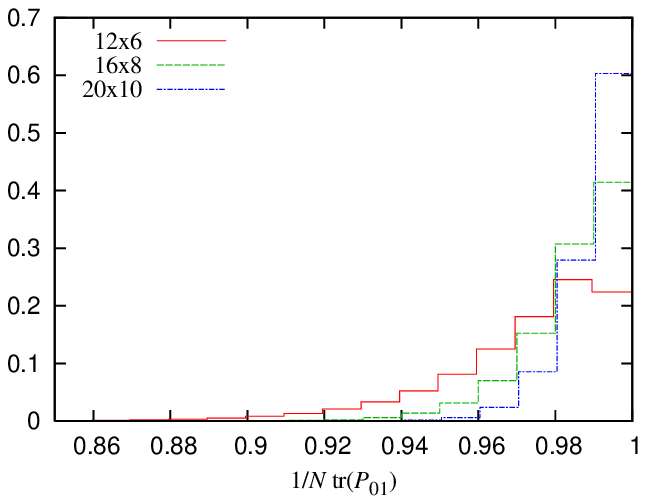}}
 \caption{ Histogram of the plaquettes for $m^2/\lambda=0.25$.
 This includes all plaquettes of every configuration. 
 As the lattice spacing approaches zero, the distribution is strongly localized at $P_{01}=1$. 
  }
 \label{fig:histp01}
\vspace{05 mm}
 {\includegraphics[width=0.85\columnwidth]{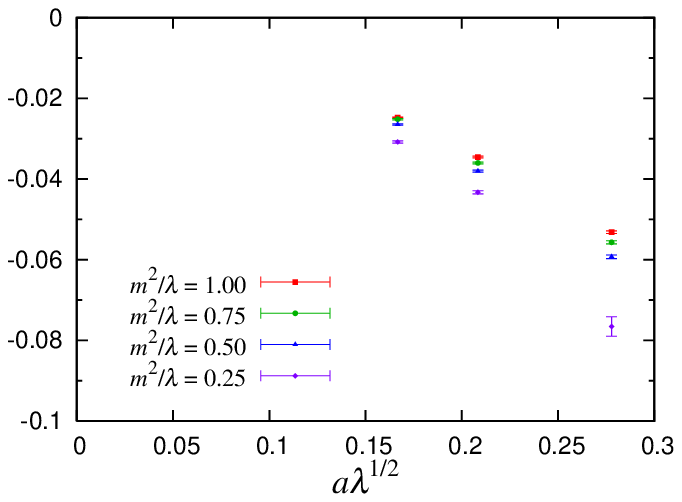}}
 \caption{ Higher derivative terms in the bosonic sector $C$ against the lattice spacing.  
 The vertical axis denotes $C/S'_B$ where $S'_B$ is the bosonic part of the action without the auxiliary fields $\sigma_\rho$. 
  }
 \label{fig:cross}
 \vspace{-05 mm}
\end{center}
\end{figure}

In this section, we show the numerical results of the SUSY WTI for $SU(2)$ and at the dimensionless temperature $T_{\rm eff}={T/\lambda^{1/2}}=0.3$.  
At this temperature, for $SU(2)$, the problem of the flat directions, which was explained in section  \ref{subsec:simulations}, occurs. Therefore, we added the mass term (\ref{mass_term}) to the action (\ref{lat_action}).
Table \ref{tab:para} summarizes the simulation parameters. 
We used four different physical masses and three different lattice spacings.
The first $6000$-$10000$ trajectories were discarded for thermalization, after that we stored 1 configuration every 20 trajectories.
The other details of the simulations are given in  section \ref{subsec:simulations}.

\begin{figure}[h!]
\begin{center}
 {\includegraphics[width=0.4\columnwidth,clip]{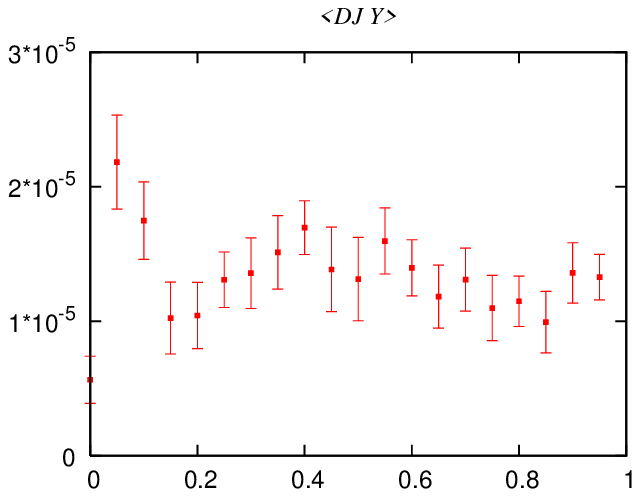}}
  \hspace{20mm}
 {\includegraphics[width=0.4\columnwidth,clip]{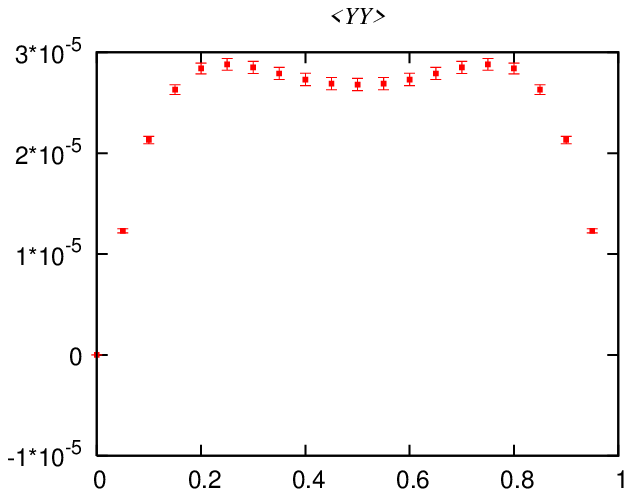}}
\end{center}
\begin{center}
 {\includegraphics[width=0.4\columnwidth,clip]{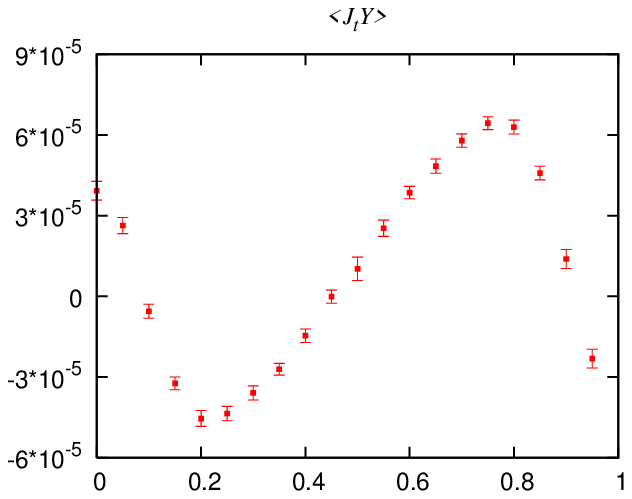}}
  \hspace{20mm}
 {\includegraphics[width=0.4\columnwidth,clip]{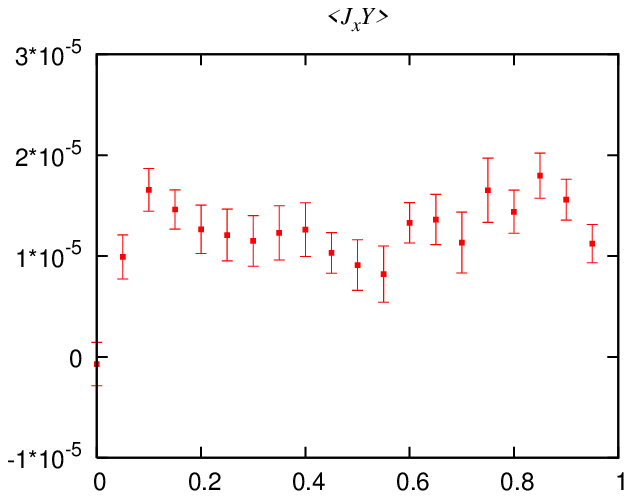}}
\end{center}
\caption{Correlation functions with respect to the WTI.
 The corresponding parameters are $N_t=20$, $m^2/\lambda = 0.5$.
 All the correlators are presented as a function of $t/N_t$ at fixed
$x=N_x/2$ for $\alpha =\beta= 1$.
The other correlators for $\alpha=\beta=2,\cdots,16$ behave in the same way.
}
\label{fig:opperator}
\end{figure}

As explained in section \ref{subsec:many_vacua},  the lattice action has many extra vacua.
The first interesting result is a confirmation that the presence of the vacua is not an issue in practice.
For $SU(2)$, the plaquettes have two minima, $+1$ and $-1$. The first one is physical, while the second one is a lattice artifact.
As can be seen in the histogram  of ${\rm tr}(P_{01})/N$ (figure \ref{fig:histp01}), the plaquettes stay close to one, we never observed any negative values in the present parameters set. 
Moreover, figure \ref{fig:cross} shows the contributions of the higher derivative terms in the bosonic sector, $C$ (\ref{full_c_term}).  
For all parameters, $C$ is relatively small, only a few percent of the action, and approaches zero as $a\rightarrow 0$.
The presence of the mass plays an important role in suppressing $C$ because the suppressions become larger as the mass increases.
Thus, we can conclude that the extra vacua do not affect our main results given below.

To investigate the restoration of supersymmetry, we compare the lattice counterparts of the ratios (\ref{eq:ra}) with the expected value $m^2/\lambda$.
In the present notation of the gamma matrices, given in appendix \ref{def:twisted_fields},
the diagonal parts $\alpha = \beta$ show the strongest signal and are used for our analysis.
We thus have sixteen ratios corresponding to the sixteen supercharges, where all the ratios should be equivalent under the  $O(8)$ internal symmetry in the continuum limit.
The inversion of the Dirac operator is computed for all points-to-all points using LAPACK\cite{lapack}.  
The correlation functions are averaged over the coordinates to reduce the statistical errors.  
Figure \ref{fig:opperator} shows an example of the shape of the correlation
functions.

\begin{figure}[t]
\begin{center}
{\includegraphics[width=0.85\columnwidth]{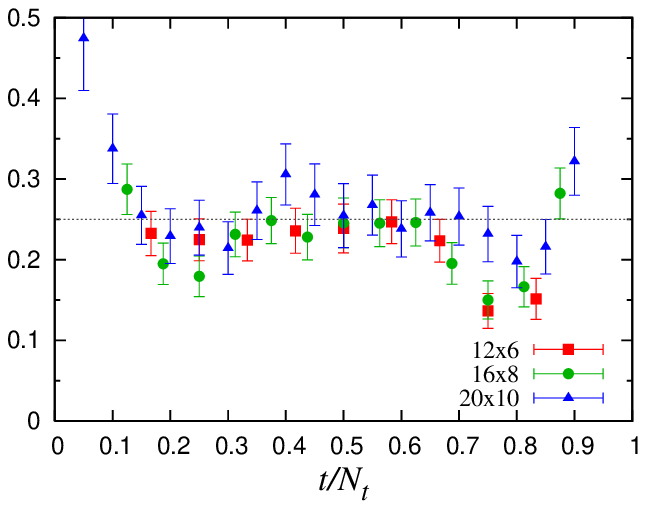}}
\caption{$\frac{\langle \partial_\mu J_{\mu 1}(0) Y_1(t,x)\rangle}{ \langle Y_1(0)Y_1(t,x)\rangle}$ as a function of $t/N_t$ at fixed  $x=N_x/2$ for $m^2/\lambda=0.25$ (denoted by the dashed line). }
\label{fig:r_cut}
\end{center}
\vspace{10mm}
\end{figure}

Figure \ref{fig:r_cut} shows the ratio for $\alpha=\beta=1$ with $m^2/\lambda=0.25$ against $t$ at fixed $x=N_x/2$.
The errors  are estimated using the jackknife analysis.
We see that in the middle range of the plot, the signal is relatively a constant and in good accordance with $m^2/\lambda$.
Close to the borders ($t=0,N_t$), we can see deviations from the constant, which correspond to the contact term of (\ref{SWTI_as_correlator}).
To quantify the recovery of the SUSY WTI,
we estimate the value of the plateau for each $\alpha$ by performing a constant fit in a rectangular area in which the plateau can be clearly seen.
The choice of the fit areas is irrelevant to the results as far as $\chi^2/dof \simeq 1$.

\begin{table}
\begin{center}
\begin{tabular}{| c c | c c c |}
  \hline 
\ \ lattice size & $\ \ \ m^2/\lambda\ \ \ $   &  fit value & $\chi^2/N_{dof}$   & fit range \\
  \hline 
  $12 \times 6$  & $1.00$  & \ \ \ 1.007 $\pm$ 0.043\  & 0.51  & $3  \times 5$ \\
  $12 \times 6$  & $0.75$  & \ \ \ 0.750 $\pm$ 0.029\  & 0.59  & $3  \times 5$ \\
  $12 \times 6$  & $0.50$  & \ \ \ 0.502 $\pm$ 0.013\  & 1.42  & $5  \times 5$ \\
  $12 \times 6$  & $0.25$  & \ \ \ 0.254 $\pm$ 0.006\  & 1.10  & $5  \times 5$ \\
  \hline
  $16 \times 8$  &  1.00   & \ \ \ 0.990 $\pm$ 0.044\  & 0.59  & $5  \times 5$ \\  
  $16 \times 8$  &  0.75   & \ \ \ 0.714 $\pm$ 0.029\  & 0.48  & $5  \times 5$ \\  
  $16 \times 8$  &  0.50   & \ \ \ 0.513 $\pm$ 0.015\  & 0.60  & $5  \times 7$ \\  
  $16 \times 8$  &  0.25   & \ \ \ 0.249 $\pm$ 0.005\  & 0.93  & $7  \times 7$ \\  
  \hline
  $20 \times 10$ &  1.00   & \ \ \ 1.001 $\pm$ 0.043\  & 0.48  & $9  \times 5$ \\
  $20 \times 10$ &  0.75   & \ \ \ 0.745 $\pm$ 0.026\  & 0.69  & $9  \times 7$ \\
  $20 \times 10$ &  0.50   & \ \ \ 0.482 $\pm$ 0.012\  & 1.03  & $11 \times 7$ \\
  $20 \times 10$ &  0.25   & \ \ \ 0.247 $\pm$ 0.005\  & 0.93  & $11 \times 7$ \\
  \hline  
\end{tabular}
\caption{Values of the plateau averaged over $\alpha$.}
  \label{tab:chi2} 
\end{center}
\vspace{15mm}
\end{table}

\begin{figure}[h!]
\begin{center}
 {\includegraphics[width=0.80\columnwidth]{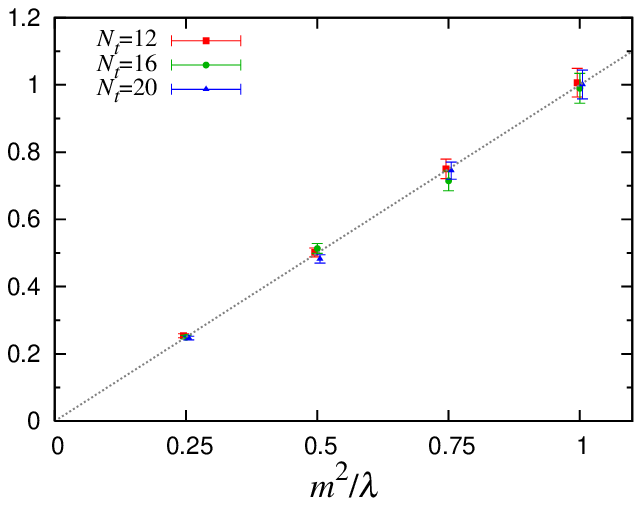}}
 \caption{ Values of the plateau averaged over $\alpha$ against the SUSY breaking mass $m^2/\lambda$. 
  All the values are on the dashed line passing through the origin with a slope of one. 
            }
 \label{fig:m_2_0}
 \vspace{15 mm}
%
 {\includegraphics[width=0.80\columnwidth]{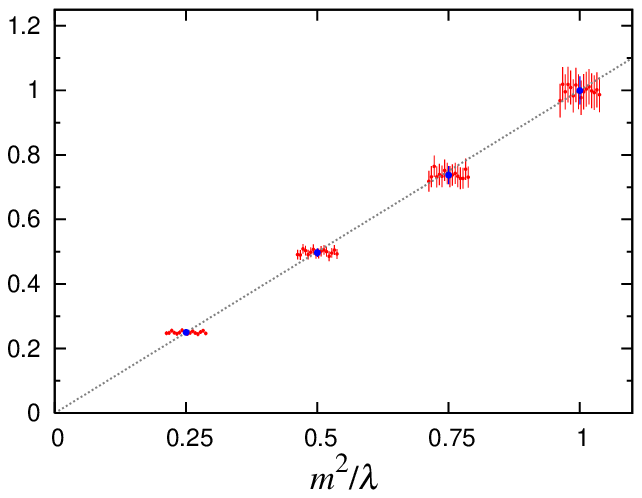}}
 \caption{ Continuum limit of the values of the plateau obtained for each spinor. The $x$-axis denotes $m^2/\lambda$.
 The red points with errors separated by a little distance denote the sixteen values corresponding spinors $\alpha=1,\cdots,16$ from left to right. 
The blue points denote the averaged values.
All the values are on the dashed line passing through the origin with a slope of one. 
 }
 \label{fig:to_0}
\end{center}
\vspace{-10 mm}
\end{figure}
Table \ref{tab:chi2} shows the values averaged over  $\alpha$ with $\chi^2$ and the fit ranges. 
We find that the fitted values of the plateaus show good agreement with $m^2/\lambda$. 
We also show the results as figure \ref{fig:m_2_0}.
Clearly,  all the values are on a straight line passing through the origin with a slope of one. 
In the massless limit, our results indicate that the supercurrent is conserved as expected.
These results tell us that the SUSY WTI holds if we ignore the contact term.

\begin{table}
\begin{center}
\begin{tabular}{| c c | c  |}
  \hline 
\ \ lattice size & $\ \ \ m^2/\lambda\ \ \ $   &  $N/V$ \\
  \hline 
   && \vspace{-4mm} \\
  $12 \times 6$  & 1.00  &  $\ \ \ 0.50 \ \pm \ 0.03 \ \ \, {}^{+0.07}_{-0.06} \ \  \ $ \\
  && \vspace{-4mm} \\
  $12 \times 6$  & 0.75  &  $\ \ \ 0.49 \ \pm \ 0.05 \ \ \, {}^{+0.05}_{-0.03} \ \ \ $  \\
   && \vspace{-4mm} \\
  $12 \times 6$  & 0.50  &  $\ \ \ 0.49 \ \pm \ 0.06 \ \ \, {}^{+0.05}_{-0.06} \ \ \ $  \\
   && \vspace{-4mm} \\
  $12 \times 6$  & 0.25  &  $\ \ \ 0.47 \ \pm \ 0.07 \ \ \, {}^{+0.05}_{-0.06} \ \ \ $  \\
   && \vspace{-4mm} \\
  \hline
   && \vspace{-4mm} \\
  $16 \times 8$  & 1.00  &  $\ \ \ 0.57 \ \pm \ 0.05 \ \ \, {}^{+0.03}_{-0.06} \ \ \ $  \\
   && \vspace{-4mm} \\
  $16 \times 8$  & 0.75  &  $\ \ \ 0.56 \ \pm \ 0.07 \ \ \, {}^{+0.05}_{-0.04} \ \ \ $  \\
   && \vspace{-4mm} \\
  $16 \times 8$  & 0.50  &  $\ \ \ 0.56 \ \pm \ 0.05 \ \ \, {}^{+0.05}_{-0.05} \  \ \  $  \\ 
   && \vspace{-4mm} \\
  $16 \times 8$  & 0.25  &  $\ \ \ 0.53 \ \pm \ 0.05 \ \ \, {}^{+0.05}_{-0.04} \ \ \ $  \\ 
   && \vspace{-4mm} \\
  \hline
   && \vspace{-4mm} \\
  $20 \times 10$ & 1.00  &  $\ \ \ 0.64 \ \pm \ 0.04 \ \ \, {}^{+0.04}_{-0.05} \ \ \  $  \\
   && \vspace{-4mm} \\
  $20 \times 10$ & 0.75  &  $\ \ \ 0.65 \ \pm \ 0.04 \ \ \, {}^{+0.03}_{-0.03} \ \ \  $  \\
   && \vspace{-4mm} \\
  $20 \times 10$ & 0.50  &  $\ \ \ 0.65 \ \pm \ 0.04 \ \ \, {}^{+0.03}_{-0.02} \ \ \ $  \\
   && \vspace{-4mm} \\
  $20 \times 10$ & 0.25  &  $\ \ \ 0.63 \ \pm \: 0.03 \ \ \, {}^{+0.04}_{-0.03} \ \ \  $  \\
  \hline  
\end{tabular}
\caption{$N/V$ for each parameters set.  The first and the second column in $N/V$ denote $N/V$ averaged over $\alpha$ and their statistical errors, respectively. While the third column denotes differences among spinors. The upper and lower values in the third column represent the maximum and minimum values of the differences between each $N_\alpha/V$ and the averages, respectively.}
  \label{tab:vol_2_c} 
\end{center}
\end{table}

In the continuum limit, the ratios should be  $m^2/\lambda$ everywhere except at the origin.
On the lattice, they behave as constants only far from the origin because the contact terms of (\ref{SWTI_as_correlator}) are smeared.
To investigate the behavior of the smeared contact terms, let us count the numbers of sites on which the ratios are $m^2/\lambda$ within the errors.
To do this, for subsets of the lattice points $\Gamma_\alpha$,
we define 
\begin{eqnarray}
\chi_\alpha^2(\Gamma_\alpha) = \sum_{\vec{x}\in \Gamma_\alpha} 
\frac{ R_\alpha^2(\vec{x}) }{ ( {\rm error \ of \  }  R_\alpha(\vec{x}) )^2}, 
\end{eqnarray}
where
\begin{eqnarray}
R_\alpha(\vec{x}) = \frac{  \partial_\mu \langle J_{\mu, \alpha} (0) Y_\alpha(\vec{x})\rangle}{ \langle Y_\alpha(0)Y_\alpha(\vec{x})\rangle} - \frac{m^2}{\lambda}.
\end{eqnarray}
We took the largest possible $\Gamma_\alpha$ within $\chi_\alpha^2(\Gamma_\alpha)/N_\alpha \le 1$ , where $N_\alpha$ are the numbers of sites in $\Gamma_\alpha$. Roughly speaking, $\Gamma_\alpha$ are the plateau areas.
In the continuum limit, the smeared contact terms should become the genuine contact terms.
Hence, the ratios $N_\alpha/V$ must approach one as the lattice spacing decreases.
Table \ref{tab:vol_2_c} shows the $N_\alpha/V$ ratios.
We find that $N_\alpha/V$ do indeed increase as the lattice spacing decreases.
This means that the smeared contact terms tend toward the expected delta function in the continuum limit.

Finally, we study whether the same results can be obtained for each supercharge.
We have already seen that  the SUSY WTI averaged over $\alpha$ holds in the continuum limit.
All we have to do is to see the degeneracy of the results.
Figure  \ref{fig:to_0} shows  the fitted values of the plateau for all the sixteen supercharges.
For the extrapolation to the continuum limit, we use constant fits. 
The values coincide with each other within the statistical errors.
Namely, the ratios are the same as $m^2/\lambda$ independently of the spinors.
Moreover, $N_\alpha/V$ in table  \ref{tab:vol_2_c} suggest that the contact terms approach the delta function for each spinor.
Therefore, supersymmetries are recovered for all the spinor indices.

%
%
%
%
%

\section{Summary and discussion}
\label{sec:summary}

We are performing lattice simulations of two-dimensional SYM with sixteen supercharges to verify the duality conjecture.
When starting from the lattice action preserving two of the sixteen supercharges on the lattice, a non-trivial and crucial question is whether 
all the sixteen supercharges are restored in the continuum limit. 
The numerical results of the SUSY WTI tells us that 
the SUSY breaking cut-off effect vanishes and the remaining fourteen supercharges are restored in the continuum limit.
This is the first result showing that the Sugino action does work for two-dimensional ${\cal N}=(8,8)$ SYM.

The lattice ${\cal N}=(8,8)$ SYM has many degenerate vacua. 
However, we have not observed the degeneracy in the simulations with the SUSY breaking mass term in $SU(2)$ presented in this paper, nor in subsequent simulations at high temperatures without the mass term in $SU(8)$.
When we start the simulations from the trivial configuration ($U_\mu=1, X_i=0$),  at high temperature $T/\lambda^{1/2} \gg 1$, both temporal and spatial Wilson loops and the plaquettes tend to stay close to unity,  and the scalar fields distribute around zero.
In other words, transitions from trivial vacuum to the others would be suppressed at high temperature, for sufficiently large $N$, or near the continuum limit so that our choice of the lattice field tensor (\ref{F01}) is justified.  
The transitions were, somewhat surprisingly, not observed at low temperature such as $T/\lambda^{1/2}=0.3$ in $N=2$ and at the lattice spacing $a\lambda^{1/2} \simeq 0.28$ at which the SUSY WTI were measured.  
The degenerate vacua might not be a problem for a wide range of simulation parameters.

The next step of this study is to investigate the phase structure and to test the duality conjecture\cite{Itzhaki:1998dd}.
To see the correspondence between the gauge theory and the black 1-branes, 
we have to choose simulation parameters in which the theory is well-described by the black 1-branes.
Then, from lattice simulations {with the determined parameters,  we will estimate the internal energy of the branes\cite{Klebanov:1996un}.  
To do such verification of the duality, further simulations are ongoing.

%
%
\section*{Acknowledgment}
We would like to thank N.Kawamoto for encouraging us to do this project.
D.K. also thank the system engineers of RICC for hospitality, as well as the staff of HPCI helpdesk (KEI computer).
The numerical calculations were performed by using the RIKEN Integrated Cluster of Clusters (RICC) facility,
RIKEN's K computer, KEK supercomputer and SR16000 at  YITP in Kyoto University.

\appendix

%
%
%
%
%

\section{Definition of twisted fields}
\label{def:twisted_fields}
We use two different notations of the fields in this paper. 
The action (\ref{cont_action}) is written in the original variables $A_\mu, X_i, \Psi_\alpha$ for $\mu=0,1,\ i=2,\cdots,9, \  \alpha=1,\cdots,16$, 
while the twisted action (\ref{exact_cont_action}) is written in the twisted variables,  $A_\mu,B_i,C, \phi_{\pm}$ and $\eta_{\pm}, \psi_{\pm\mu}, \chi_{\pm i}$ for $\mu=0,1,2,3$ and $i=0,1,2$.
In this appendix, we give the  relation between the two notations.

Let us express the action (\ref{cont_action}) written in  $A_\mu, X_i, \Psi_\alpha$ as the $Q_\pm$-exact action  (\ref{exact_cont_action}) written in the twisted variables $A_\mu,B_i,C, \phi_{\pm}$.
The gauge fields are unchanged,  and the other bosonic twisted fields $A_2,A_3,B_i,C, \phi_{\pm}$ are given as follows:
\begin{eqnarray}
&& A_\mu = X_\mu, \qquad (\mu=2,3),
\nonumber
\\
&& B_0=-X_6, 
\nonumber
\\
&&  B_1=X_5, 
\nonumber
\\
&&  B_2=- X_4,
 \label{def:X_to_twisted_boson}
\\
&& C=2 X_7, 
\nonumber
\\
&& \phi_+ = X_8 -i X_9,
\nonumber
\\
&& \phi_- = -X_8 -i X_9.
\nonumber
\end{eqnarray}
The fermionic twisted variables $\eta_{\pm}, \psi_{\pm\mu}, \chi_{\pm i}$ and $\Psi_\alpha $ can be related to each other by
\begin{eqnarray}
&&
\left( \hspace{-1mm}
  \begin{array}{c}
     \Psi_1 \\
     \Psi_2 \\
     \Psi_3 \\
     \Psi_4 \vspace{-4mm}\\
                   \\
     \Psi_5 \\
     \Psi_6 \\
     \Psi_7 \\
     \Psi_8 \vspace{-4mm} \\
                   \\       
     \Psi_{9} \\
     \Psi_{10} \\
     \Psi_{11} \\
     \Psi_{12} \vspace{-4mm}\\
                   \\
     \Psi_{13} \\
     \Psi_{14} \\
     \Psi_{15} \\
     \Psi_{16} \\
  \end{array}
\hspace{-1mm} \right)
= \frac{1}{\sqrt{2}}
\left( \hspace{-1mm}
  \begin{array}{c}
     \psi_{-0}   + \frac{i}{2} \eta_{+} \\
     \psi_{-1}   - i           \chi_{+2} \\
     \psi_{-2}   + i           \chi_{+1} \\
     \psi_{-3}   - i           \chi_{+0} \vspace{-4mm} \\
     \\
     \psi_{+0}   + \frac{i}{2} \eta_{-} \\
     \psi_{+1}   + i           \chi_{-2} \\
     \psi_{+2}   - i           \chi_{-1} \\
     \psi_{+3}   + i           \chi_{-0} \vspace{-4mm} \\
     \\
     -i(\psi_{-0}   - \frac{i}{2} \eta_{+} ) \\
     -i(\psi_{-1}   + i           \chi_{+2}) \\
     -i(\psi_{-2}   - i           \chi_{+1}) \\
     -i(\psi_{-3}   + i           \chi_{+0}) \vspace{-2mm} \\
     \\
     -i(\psi_{+0}   - \frac{i}{2} \eta_{-}) \\
     -i(\psi_{+1}   - i           \chi_{-2}) \\
     -i(\psi_{+2}   + i           \chi_{-1}) \\
     -i(\psi_{+3}   - i           \chi_{-0}) 
  \end{array}
\hspace{-1mm} \right),
\label{def:psi_to_twisted_fermion}
\end{eqnarray}
with the following representation of gamma matrices,
\begin{eqnarray}
&& \gamma_1 = \sigma_2  \otimes  \, {\mathbf 1} \,  \otimes  \sigma_3          \otimes  \sigma_2,           
\nonumber \\
&& \gamma_2 = \sigma_2  \otimes  \, {\mathbf 1} \,  \otimes  \sigma_2          \otimes  \, {\mathbf 1} \,,  
\nonumber \\
&& \gamma_3 = \sigma_2  \otimes  \, {\mathbf 1} \,  \otimes  \sigma_1          \otimes  \sigma_2,           
\nonumber \\
&& \gamma_4 = \sigma_3  \otimes  \sigma_2           \otimes  \, {\mathbf 1} \, \otimes  \sigma_2,           
\nonumber \\
&& \gamma_5 = \sigma_3  \otimes  \sigma_2           \otimes  \sigma_2          \otimes  \sigma_3,           
\label{def:nine-gamma-matrices} \\
&& \gamma_6 = \sigma_3  \otimes  \sigma_2           \otimes  \sigma_2          \otimes  \sigma_1,           
\nonumber \\
&& \gamma_7 = \sigma_3  \otimes  \sigma_1           \otimes  \, {\mathbf 1} \, \otimes  \, {\mathbf 1} \, , 
\nonumber \\
&& \gamma_8 = \sigma_3  \otimes  \sigma_3           \otimes  \, {\mathbf 1} \, \otimes  \, {\mathbf 1} \, , 
\nonumber \\
&& \gamma_9 = \sigma_1  \otimes  \, {\mathbf 1} \,  \otimes  \, {\mathbf 1} \, \otimes  \, {\mathbf 1},
\nonumber
\end{eqnarray}
where $\sigma_i$ are the Pauli matrices. 
\footnote{
The gamma matrices act on the fermions $\psi_\alpha$
as $(\gamma_1)_{\alpha_1\alpha_2} \Psi_{\alpha_2}$ where the spinor index $\alpha$ corresponds to the indices of the Pauli matrices as follows: 
\begin{eqnarray}
(\gamma_1)_{\alpha_1\alpha_2} =(\sigma_2)_{i_1i_2}  
\otimes {\mathbf 1}_{j_1j_2} \otimes  (\sigma_3)_{k_1k_2} \otimes (\sigma_2)_{l_1l_2}, 
\end{eqnarray}
with $\alpha=8(i-1) +4(j-1)+2(k-1) +l$.
}

The two scalar supercharges $Q_\pm$ are defined as
\begin{eqnarray}
&& Q_+ = \frac{1}{\sqrt{2}}(Q_5 - i Q_{13}),\\
&& Q_- = \frac{1}{\sqrt{2}}(Q_1 - i Q_{9}).
\end{eqnarray}
The $Q_\pm$-transformations of the fields come from those of $Q_\alpha$ given in (\ref{Q_cont_A}),  (\ref{Q_cont_X}) and  (\ref{Q_cont_psi}).

Using the definitions of the twisted fields above and the $Q_\pm$-transformations (\ref{def:Q-transf}), 
and adding the Gaussian integrals of the auxiliary fields $\tilde H_\mu$ and $H_i$ to the action (\ref{cont_action}), 
we can rewrite the action (\ref{cont_action}) as the $Q_\pm$-exact form (\ref{exact_cont_action}).

%
%
%
%
%

\section{Lattice action written in twisted variables}

Here we give the full lattice action by performing the $Q_\pm$-transformations (\ref{Lat_Q}) in the lattice $Q_\pm$-exact action (\ref{lat_action}).
First, let us introduce some useful notations and covariant difference operators.
Then, by using them, we present the lattice boson and fermion actions.

The following shifts
\begin{align}
&\varphi^{+\mu}(\vec{x}) =  U_\mu(\vec{x})\varphi(\vec{x}+\hat{\mu})U_\mu^\dagger(\vec{x}),  & (\mu=0,1),    \label{fwd_shift}\\
&\varphi^{-\mu}(\vec{x}) =  U_\mu^\dagger(\vec{x}-\hat{\mu})\varphi(\vec{x}-\hat{\mu})U_\mu(\vec{x}-\hat{\mu}),  &(\mu=0,1),
  \label{bwd_shift}\\
&\varphi^{\pm \mu}(\vec{x}) =  \varphi(\vec{x}),          &(\mu=2,3) \,
\end{align}
and covariant difference operators
\begin{align}
&\nabla^{\pm}_\mu \varphi = \pm( \varphi^{\pm \mu}-\varphi ), & (\mu=0,1), \\
&\nabla^\pm_\mu \varphi = i[A_\mu,\varphi],                         & (\mu=2,3) \,
\end{align}
are useful to write the full lattice action.
Here,  the sites on which the fields are defined were abbreviated.  This is possible because the gauge covariance always tells us where the fields live. 
The lattice action has two complex covariant difference operators,
\begin{eqnarray}
&& \nabla^{+\nu}_\mu  \varphi  = \frac{1}{2} (\varphi^{+\mu} P_{\mu \nu } + P_{\nu \mu}\varphi^{+\mu }- \varphi P_{\nu \mu}    -P_{\mu \nu} \varphi  ), \\
&&\nabla^{-\nu}_\mu \varphi  = \frac{1}{2} (P_{\nu \mu}\varphi +\varphi P_{\mu \nu } -P_{\mu \nu }^{-\mu} \varphi^{-\mu}-\varphi^{-\mu}P^{-\mu}_{\nu \mu}), 
\end{eqnarray}
for $\mu,\nu=0,1,\mu\neq\nu$.

After performing the lattice $Q_\pm$-transformations (\ref{Lat_Q}), we obtain the full lattice boson action,
\begin{eqnarray}
&& S_B =\frac{N}{2\lambda_0}\sum_{t,x}  {\rm tr}\bigg \{
\frac{1}{4}[\phi_+,\phi_-]^2+\frac{1}{4}[C,\phi_+][C,\phi_-]
-\frac{1}{4}[C,B_i]^2\nonumber \\
&& \hspace{3cm} -\nabla^+_\mu \phi_+ \nabla^+_\mu \phi_- +[B_i,\phi_+][B_i,\phi_-]+ \frac{1}{4}(\nabla^+_\mu C)^2 \\
&&  \hspace{3cm} +(H_i+i\varphi_i)^2 + \varphi_i^2 +(\tilde{H}_\mu+iG_\mu)^2 + G_\mu^2 \ \  \bigg\},\nonumber 
\end{eqnarray}
where $\varphi_i$ and $G_\mu$ are given by
\begin{eqnarray}
&&\varphi_0 = \nabla^+_0   A_3 + \nabla^+_1   A_2 -i[B_1,B_2],   \\
&&\varphi_1 = \nabla^-_1 A_3 - \nabla^-_0 A_2 -i[B_2,B_0],       \\
&&\varphi_2 = i[A_2 ,A_3 ]   + F_{01}         -i[B_0,B_1], 
\end{eqnarray}
and
\begin{eqnarray}
&&G_0 =   i[A_3^{+0}, B_0] - i[A_2, B_1^{+0}] + \nabla^{-0}_1 B_2,    \\
&&G_1 =   i[A_2^{+1}, B_0] + i[A_3, B_1^{+1}] - \nabla^{-1}_0 B_2,    \\
&&G_2 = - \nabla^{-}_1 B_0 + \nabla^{+}_0 B_1 + i[A_3,B_2],          \\
&&G_3 = - \nabla^{-}_0 B_0 - \nabla^{+}_1 B_1 - i[A_2, B_2].
\end{eqnarray}

The lattice boson action is semi positive definite. 
The actions of the seven auxiliary fields $\tilde H_\mu$ and $H_i$ are given by the Gaussian integrals which can be analytically integrated.
Every terms except for $\varphi_i$ and $G_\mu$ look like the continuum ones. 
For $\varphi_i$ and $G_\mu$, some fields are shifted. 
Since the higher derivative terms in the bosonic sector,
\begin{eqnarray}
\begin{split}
C = \frac{N}{\lambda_0} \sum_{t,x} {\rm tr} \left\{  \right.
  &iF_{01}[A_2 ,A_3 ]         - \nabla^-_0 A_2  \nabla^-_1 A_3    + \nabla^+_0 A_3  \nabla^+_1   A_2  \\[-12pt]
- &iF_{01}[B_0,B_1]           + \nabla^{-}_0 B_0\nabla^{+}_1 B_1  - \nabla^{+}_0 B_1\nabla^{-}_1 B_0  \\
- &i\nabla^-_0 A_2[B_0,B_2]   +i\nabla^{-}_0 B_0[A_2,B_2]         -i\nabla^{-1}_0 B_2[A_2^{+1}, B_0]  \\
- &i\nabla^+_0 A_3[B_1,B_2]   +i\nabla^{+}_0 B_1[A_3,B_2]         -i\nabla^{-1}_0 B_2[A_3, B_1^{+1}]  \\
- &i\nabla^+_1 A_2[B_1,B_2]   +i\nabla^{+}_1 B_1[A_2,B_2]         -i\nabla^{-0}_1 B_2[A_2, B_1^{+0}]  \\
+ &i\nabla^-_1 A_3[B_0,B_2]   -i\nabla^{-}_1 B_0[A_3,B_2]         +i\nabla^{-0}_1 B_2[A_3^{+0}, B_0]  \\
+ & \left.[A_2 ,A_3 ] [B_0,B_1]      - [A_2^{+1}, B_0][A_3, B_1^{+1}]    + [A_2, B_1^{+0}][A_3^{+0}, B_0] \right \},
\end{split}
\label{full_c_term}
\end{eqnarray}
disappear in the naive continuum limit,   
the lattice boson action reproduces the continuum boson action in (\ref{cont_action}) by using (\ref{def:X_to_twisted_boson}).

The fermion action is given by
\begin{align}
\begin{split}
S_F = \frac{N}{2\lambda_0} \sum_{t,x}
  {\rm tr} \bigg \{  & \  \frac{1}{4}\eta_+[\phi_-,\eta_+] +\chi_{+i}[\phi_-,\chi_{+i}] +\psi_{+\mu}[ \tfrac{1}{2}(\phi_- + \phi_-^{+\mu}) ,\psi_{+\mu}] \\
&+\frac{1}{4}\eta_-[\phi_+,\eta_-] +\chi_{-i}[\phi_+,\chi_{-i}] +\psi_{-\mu}[ \tfrac{1}{2}(\phi_+ + \phi_+^{+\mu}) ,\psi_{-\mu}] \\
&-\frac{1}{4}\eta_+[  C   ,\eta_-] +\chi_{+i}[  C   ,\chi_{-i}] +\psi_{+\mu}[ \tfrac{1}{2}(     C +      C^{+\mu}) ,\psi_{-\mu}] \\
&-\eta_-[ B_i , \chi_{+i} ]-\eta_+[ B_i , \chi_{-i} ]+2\epsilon_{ijk}\chi_{-i}[B_j,\chi_{+k}]\\
&-i \nabla_\mu \eta_+ \psi_{-\mu}-i \nabla_\mu \eta_- \psi_{+\mu}+ [A_\mu, \eta_+] \psi_{-\mu} + [A_\mu, \eta_- ]\psi_{+\mu}\\
&-2\chi_{+0}( +i\nabla^+_0 \psi_{-3}  +i\nabla^+_1 \psi_{-2} +[A_2^{+1},\psi_{-1}]      +[A_3^{+0},\psi_{-0}] )\\
&+2\chi_{-0}( +i\nabla^+_0 \psi_{+3}  +i\nabla^+_1 \psi_{+2} +[A_2^{+1},\psi_{+1}]      +[A_3^{+0},\psi_{+0}] )\\
&-2\chi_{+1}( -i\nabla^-_0 \psi_{-2}  +i\nabla^-_1 \psi_{-3} -[A_2^{-0},\psi^{-0}_{-0}] +[A_3^{-1},\psi^{-1}_{-1}] )\\
&+2\chi_{-1}( -i\nabla^-_0 \psi_{+2}  +i\nabla^-_1 \psi_{+3} -[A_2^{-0},\psi^{-0}_{+0}] +[A_3^{-1},\psi^{-1}_{+1}] )\\
&-2\chi_{+2}( +i\nabla^{+1}_0 \psi_{-1} -i\nabla^{+0}_1 \psi_{-0}          -[A_2, \psi_{-3}]          +[A_3, \psi_{-2}] )\\
&+2\chi_{-2}( +i\nabla^{+1}_0 \psi_{+1} -i\nabla^{+0}_1 \psi_{+0}          -[A_2, \psi_{+3}]          +[A_3, \psi_{+2}] )\\
&-2\psi_{-3}[B_0^{-0},\psi^{-0}_{+0}] -2\psi_{-2}[B_0^{-1},\psi_{+1}^{-1}]\\
&-2\psi_{-3}[B_1^{+1},\psi_{+1}]      +2\psi_{-2}[B_1^{+0},\psi_{+0}]\\
&+2\psi_{+3}[B_0^{-0},\psi^{-0}_{-0}] +2\psi_{+2}[B_0^{-1},\psi_{-1}^{-1}]\\
&+2\psi_{+3}[B_1^{+1},\psi_{-1}]      -2\psi_{+2}[B_1^{+0},\psi_{-0}]\\
&-2\psi_{-3}[B_2,\psi_{+2}]+2\psi_{+3}[B_2,\psi_{-2}]\\
&-i\psi_{-0}   [ A_3^{+0},  [ B_0 , \psi_{+0}] ]  
 -i\psi_{-0}   [B_0,  [ A_3^{+0} , \psi_{+0} ]] \\
&-i\psi_{-0}   [[A_2,   [ B_1^{+0}, \psi_{+0}] ]  
 -i\psi_{-0}   [B_1^{+0},  [ A_2 , \psi_{+0} ]] \\
&+i\psi_{-1}   [A_3,   [B_1^{+1}  , \psi_{+1}] ]  
 +i\psi_{-1}   [B_1^{+1},  [ A_3 , \psi_{+1} ]] \\
&-i\psi_{-1}   [A_2^{+1},   [B_0 , \psi_{+1}] ]  
 -i\psi_{-1}   [B_0,  [ A_2^{+1} , \psi_{+1} ]] \\
& + {\cal L}_{P}\\
&-\frac{1}{4} \sum_{\rho=0,1}\{\psi_{+\rho},\psi_{-\rho}\}^2 \ \ \bigg\},
\label{lattice_fermi_action}
\end{split}
\end{align}
where
\begin{eqnarray}
\begin{split}
{\cal L}_{P} = &  
\frac{1}{2} \psi_{-0}  \left\{ (P_{10} B_2 -B_2P_{01} )^{-1}  + B_2 P_{10} - P_{01} B_2 , \psi_{+0} \right\} \\
& +\psi_{-0}   P_{01}       \psi_{+0}^{+1}  B_2               
   -\psi_{-0}   B_2         \psi_{+0}^{+1}  P_{10}       
 +\psi_{-0}^{+1}   B_2    \psi_{+0}  P_{01}  
   -\psi_{-0}^{+1}  P_{10}  \psi_{+0}  B_2   \\
&- \frac{1}{2} \psi_{-1}  \left\{ (P_{01} B_2 -B_2P_{10} )^{-1}  + B_2 P_{01} - P_{10} B_2 , \psi_{+1} \right\} \\
& -\psi_{-1}   P_{10}       \psi_{+1}^{+0}  B_2               
   +\psi_{-1}   B_2         \psi_{+1}^{+0}  P_{01}       
 -\psi_{-1}^{+0}   B_2    \psi_{+1}  P_{10}  
   +\psi_{-1}^{+0}  P_{01}  \psi_{+1}  B_2   \\
& +\psi_{-0}  P_{01}          \psi_{+1}         B_2   
    -\psi_{-0}  B_2            \psi_{+1}         P_{10}         
 +\psi_{-0}^{+1}  P_{10}     B_2         \psi_{+1}      
    -\psi_{-0}^{+1}  \psi_{+1}  B_2         P_{01}    \\
& - \psi_{-0} \psi_{+1}^{+0}  P_{01}            B_2      
    +\psi_{-0}  B_2            P_{10}            \psi_{+1}^{+0} 
 +\psi_{-0}^{+1}  B_2       \psi_{+1}^{+0}  P_{01}     
    -\psi_{-0}^{+1}  P_{10}     \psi_{+1}^{+0}  B_2      \\
& -\psi_{-1}  P_{10}  \psi_{+0}   B_2   
    +\psi_{-1}  B_2  \psi_{+0} P_{01}     
 -\psi_{-1}^{+0} P_{01}    B_2 \psi_{+0}      
    +\psi_{-1}^{+0}  \psi_{+0} B_2 P_{10}     \\
& +\psi_{-1}  \psi_{+0}^{+1} P_{10}   B_2      
    -\psi_{-1}  B_2 P_{01} \psi_{+0}^{+1}   
 -\psi_{-1}^{+0}  B_2  \psi_{+0}^{+1} P_{10}     
     +\psi_{-1}^{+0}   P_{01} \psi_{+0}^{+1}  B_2.
\end{split}
\end{eqnarray}

In the naive continuum limit, the complicated term ${\cal L}_{P}$ becomes two simple terms,
\begin{eqnarray} 
 {\cal L}_{P}  = -2 \psi_{-1}  [B_2, \psi_{+0}] +2  \psi_{-0}  [B_2,            \psi_{+1}],
\end{eqnarray} 
while the shifts of fields in (\ref{lattice_fermi_action}) disappears. 
 Thus, by using (\ref{def:X_to_twisted_boson}), (\ref{def:psi_to_twisted_fermion}) and (\ref{def:nine-gamma-matrices}), we find that the lattice fermion action reproduces the continuum fermion action in (\ref{cont_action}).

%
%
%
%
%

\end{document}